\documentclass[conference]{IEEEtran}
\usepackage{datetime}
\usepackage{times}
\usepackage{latexsym,subfigure,makeidx}
\usepackage{epsfig,graphicx} 
\usepackage{graphicx, amssymb}
\usepackage{amsfonts, enumerate}
\usepackage{url}
\usepackage{tabls}
\usepackage{latexsym, makeidx}
\usepackage{algorithm}
\usepackage{algorithmicx}
\usepackage{algpseudocode}
\usepackage{lscape}
\usepackage[sort,space]{cite}
\usepackage{pifont}
\usepackage{tabularx}
\usepackage{rotating}
\usepackage{hhline}
\usepackage{textcomp,booktabs}
\usepackage{colortbl}
\usepackage{multirow}
\usepackage{array}
\usepackage{amsmath,bm}

\usepackage[]{caption2}
\usepackage{multirow}
\usepackage{verbatim} 

\usepackage{bm}
\usepackage[amsmath,thmmarks]{ntheorem}
\usepackage[
top    = 1.70cm,
bottom = 1.05in,
left   = 0.63 in,
right  = 0.63 in]{geometry}

\algdef{SE}[DOWHILE]{Do}{doWhile}{\algorithmicdo}[1]{\algorithmicwhile\ #1}%

\newcommand{\sr}{\text{PACO}}
\newcommand{\brick}{\text{pathlet}}
\newcommand{\bricks}{\text{pathlets}}
\newcommand{\rpath}{\text{concatenation path}}
\newcommand{\rpaths}{\text{concatenation paths}}
\newcommand{\cbrick}{\text{Pathlet}}

\newcommand{\alg}{\text{Algorithm 1}}

\newcommand{\algpath}{\text{Algorithm 3}}

\newtheorem{mytheorem}{Theorem}

\begin{document}
\title{Scalable Fine-grained Path Control in Software Defined Networks}

\ifx \drafting\undefined
\author{
\IEEEauthorblockN{Long Luo, Hongfang Yu, Shouxi Luo}

\IEEEauthorblockA{
Key Laboratory of Optical Fiber Sensing and Communications, Ministry of Education
\\University of Electronic Science and Technology of China, Chengdu, P. R. China
}
\vspace{-2ex}
}
\else

\author{
\textit{\today \ \currenttime}
}

\fi

\maketitle
\IEEEpeerreviewmaketitle
\begin{abstract}
The OpenFlow-based SDN is widely studied to better network performance through planning fine-grained paths. However, being designed to configure path hop-by-hop, it faces the scalability issue---both the flow table overhead and path setup delay are unacceptable for large-scale networks.
In this paper, we propose \sr, a framework based on Source Routing to address that problem through pre-installing few rules at network core and quickly pushing paths into the packet header at network edges. The straightforward implementation of SR is inefficient as it would incur heavy header overhead (\emph{e.g.}, Sourcey); other efficient approaches would sacrifice path flexibility (\emph{e.g.}, DEFO). To leverage SR efficiently and flexibly, \sr\ presents each path as a concatenation of \bricks\ and introduces algorithms to program \bricks\ and concatenate paths with minimum header overhead. Our extensive simulations confirm the scalability of \sr\ as it saves the flow table overhead up to $94\%$ compared with OpenFlow-SDN solutions and show that \sr\ outperforms SR-SDN solutions by supporting more than $40\%$ paths with negligible header overhead.
\end{abstract}

\section{Introduction and related work}\label{Introduction}
Fine-grained routing path control is crucial for today's network to accommodate emerging network services and manage itself \cite{2015fibbing,chen2015xpath,2015defo,2015diversity}. 
For security, Qos and traffic engineering purposes, the network operator would perform routine network management tasks including: 1) directing the suspicious traffic (may be from a denial-of-service attack) to a scrubber for inspection, 2) dispersing time-sensitive traffic flow across the low-latency path to meet the time-related Qos requirement, 3) load-balancing bulk traffic flows on under-utilized links to avoid congesting hot/heavy links. Mixing traffic flows over the same path (\emph{i.e.}, shortest path adopted in traditional intra-networks) is problematic as different traffic flows may pitted against each other, thus
the network should offer differentiated paths to adapt traffic flows to meet diverse requirements. 

The OpenFlow-based Software Defined Network (SDN) solution can achieve that goal \cite{2013b4,2013swan}, but it faces a fundamental challenge of scalability \cite{rSDN,2013scalability}.
The OpenFlow-SDN approaches achieve flexible path control through configuring hop-by-hop rule for each flow. The scalability issue of such Hop-by-Hop approach reflects in two aspects. One is the insufficient flow table space and the other is the complex path construction process as well the time overhead it incurs. The flow table limitation would prevent network installing paths for massive traffic flows of today's network \cite{2015aggregation,2016cacheflow}. What's worse, the Hop-by-Hop's uncontrollable path installation delay would prevent the agility of path control decision \cite{2015measuring,sigcomm14-dionysus}. For example, it would hinder the network from quickly finishing network updates frequently incurred by network events like TE optimization and failures \cite{sigcomm14-dionysus}. 
Although recent studies have shown that the rule compression technique can reduce the rule overhead to some extent, it is sensitive to the concrete content of rules. Moreover, compressed rules would complicate the management and update of paths.
Therefore, as the network scales up, the Hop-by-Hop solution would find it hard to deal with the increasing demand of forwarding paths.

To address the above scalability issue, a lot of work resorts to Source Routing (SR) technique to provide network \textit{infinite} paths \cite{jin2016yourdc,chen2015xpath,2015fabric,2016sdnsfc,2014exploring,2015defo,filsfils2015segment}.  
In the SR-SDN solution, the network pre-installs flow-independent forwarding rules at switches and encapsulates the entire flow path in the packet header according to the already configured rules.  When there is an ask for a new path, the controller only needs to do the path computation task and control an ingress switch to perform path encapsulation, which is easy even for the large-scale network. Clearly, this approach promises fast path construction as it eliminates the specific flow-dependent rule and operation on hop-by-hop switches.
However, there's no such thing as a free lunch---the bits overhead (used to encapsulate/store path) of packet is the expensive price paying for scalability.  
The straightforward implementation of SR is to 
directly put labels of each hop along a path into the packet header \cite{jin2016yourdc,2015fabric,2016sdnsfc,2014exploring}, which is inefficient as it would incurs heavy header/label overhead, leading to much extra bandwidth consumption.
Another implementation is to have the packet include one or several labels of middlepoints that a path must pass through \cite{2015defo,filsfils2015segment}. This approach adopts shortest paths between two middlepoints, thus it would compromise the path flexibility for lightweight label overhead or suffer from heavy label overhead for flexibility.
Others propose to pre-install all the desired paths using state compression algorithm and place a global label for each path in packets \cite{chen2015xpath}. Although the label overhead is lightweight, this approach still meets the scalability issue since it would fail to install all the paths when the condition of its compression algorithm does not be satisfied. Among the state-of-art implementations, we find none is efficient while preserving both the path flexibility and the scalability benefits SR brings. 

In this paper, we present \sr, an efficient SR-based approach to provide scalable and flexible path control in software defined networks. Following the principle of SR-SDN solution, \sr\ restricts the stateful flow rule at the network edges and maintains the stateless forwarding rule in the network core. To reduce the overhead of packet, \sr\ strategically represents each path as a concatenation of several \bricks/\textit{subpaths} \footnote{In contrast to pathlet routing \cite{2008pathlet}, we use pathlet to represent data path rather than for routing protocol design.}. Although the idea of constructing paths with existing pathlets is not new, the key novelty of this paper is proposing an SDN-SR framework together with several accompanied algorithms to implement it in an efficient and flexible way. 

To preserve the path flexibility, \sr\ explicitly programs \bricks\ and installs them in the network beforehand. Thus, the real challenges \sr\ tackles are to: 1) decide which \bricks\ should be pre-installed such that both the path flexibility and scalability are preserved, and 2) how to concatenate a path with minimum \bricks\ such that the goal of efficiency is achieved. 

\sr\ addresses the first challenge by converting it into a \brick\ selection linear programming problem and presenting an efficient \brick\ selection algorithm based on our formulated optimization model. As for the path concatenation problem, \sr\ presents an optimal algorithm based on the branch-and-bound technique and a \brick\ nesting scheme.

In summary, we make the following contributions.

\noindent\textbf{Analysis and Framework}. We disclose the limitations of prior work and show how to use \bricks\ to overcome them.
(section \ref{sec:motivation}). We propose a novel SDN-SR based framework to facilitate the scalable and flexible path control. This framework utilizes the global view of SDN control plane to determine the path and follows the forwarding principle of source routing (section \ref{sec:overview}).

\noindent\textbf{Modeling}. Given the desired paths, we formalize the \brick\ selection as a linear programming problem which seeks to compute appropriate \bricks\ to support all the paths under resource constraints.
(section \ref{sec:segment_selection}). 

\noindent\textbf{Algorithms}. We present a Lagrangian heuristic for the \brick\ selection problem and an optimal algorithm for the path concatenation problem.
(section \ref{subsec:lagrangian-heuristic},\ref{sec:path_construction}). 

\noindent\textbf{Experimental evaluation}. We evaluate our implementation of \sr\ by simulating extensive path control scenarios for realistic networks. Our results show that \sr\ hugely outperforms previous techniques in terms of scalability, efficiency and flexibility.
(section \ref{sec:evaluation}).

\section{Motivation and related work}
\label{sec:motivation}
\begin{figure*} \centering 
\subfigure[The network topology] {\label{subfig:topo} 
\includegraphics[width=0.95\columnwidth]{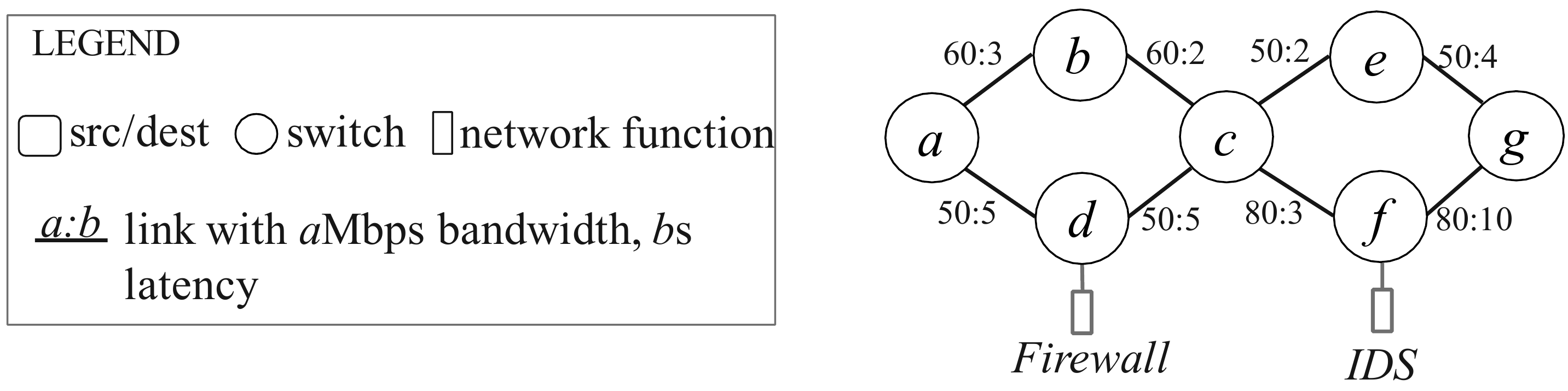} 
} 
\subfigure[Forwarding paths] {\label{subfig:path} 
\includegraphics[width=1\columnwidth]{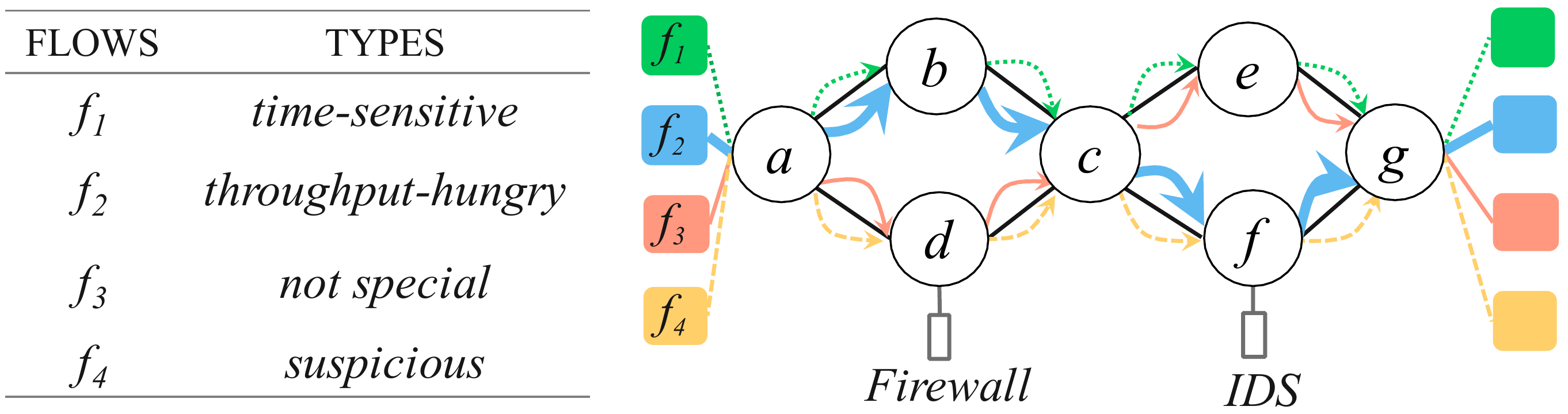} 
} 
\caption{A forwarding path map of mixed flows.} 
\label{fig:routemap} 
\vspace{-2ex}
\end{figure*}
Fig. \ref{fig:routemap} shows a scenario where the SDN controller (not depicted for brevity) need to build fine-grained routing paths for the traffic flowing across the controlled network. 
Provided that there exists four type traffic flows, posing different requirements on their own delivery, between the same ingress-egress pair (see Fig. \ref{subfig:path}). For the sake of the example, we assume that all the network state (\emph{i.e.}, link bandwidth and delay, traffic type) are acknowledged to the centralized controller (see Fig. \ref{subfig:topo}). 
Supposing the colored solid arrows represents the programmed optimal flow paths that satisfy the flow constraints (\emph{e.g.}, throughput, latency) and the network constraints (\emph{e.g.}, load balancing, secure data delivery).
 
In Fig. \ref{fig:routemap}, if the network adopts Hop-by-Hop technique to install the planned paths, it will fail when a rule-scarce node (say $c$) is capable of adding only $3$ more entries to its flow table (while the actual needed is $4$). Besides, provided the average time of installing a rule to each hop switch from the controller is $\delta$, then the waiting time before a flow enjoying its 5-hops path is $5\delta$ at worst. Thus, if the lived time of time-sensitive (say flow $f_1$) is less than $5\delta$, its data would be dropped at ingress switch before transmission.
We say a SDN solution produces a scalable, flexible path control if all the fine-grained paths can be quickly installed in the limited concrete forwarding table.

\subsection{Previous implementations have limitations}
Prior work achieves that goal by exploiting Source Routing (SR) to strictly (\emph{e.g.}, \cite{jin2016yourdc,chen2015xpath,2015fabric,2016sdnsfc,2014exploring,2015defo,filsfils2015segment}) or loosely encapsulate the path into each packet header. These SR-SDN approaches hugely mitigates the scalability issue, including flow table overhead and path setup latency, of Hop-by-Hop by moving per-flow routing state from switches to the packet header and pre-installing necessary forwarding rules. According to the encapsulation granularity, we refer current implementations as hop encapsulation (\emph{i.g.}, \cite{2014exploring,2015fabric,2016sdnsfc,jin2016yourdc}), middlepoint encapsulation (\emph{i.g.}, \cite{2015defo,filsfils2015segment}) and path encapsulation (\emph{i.g.}, \cite{chen2015xpath}) respectively. Unfortunately, they are inefficient or limited, because they only focus on the benefit but neglect the cost of SR, or sacrifice flexibility and still suffer from scalability.

\noindent\textbf{Hop encapsulation is inefficient.} It is based on strict SR by inserting labels for the entire path into the packet header at the ingress node (\cite{2014exploring,2015fabric,2016sdnsfc}) or the host side (\cite{jin2016yourdc}); each label value indicates the outgoing port number at a hop switch. Since the network diameter is always large in the networks like public WANs or carrier networks, the label overhead of the packet header would be heavy if adopted this technique. This comes with two possible consequences. First, the technique may simply not be applicable if a packet can only carry limited number of labels in the header. For example, a IPv4 packet is able to add at most 4 8-bits labels into its header because of the length constraint of the packet header (while 5 labels is asked to be added for the paths in the Fig. \ref{fig:routemap}). Second, even if multiple labels are supported (\emph{i.e.}, IPv6 extension header), inserting all the port numbers of a path on the packet header would significantly increase the packet size leading to high bandwidth consumption.

\noindent\textbf{Middlepoint encapsulation has limited flexibility.}
They are based on loose SR by putting labels of middlepoints into each packet header at ingress nodes; each middlepoint is a waypoint that the path must go through \cite{2015defo,filsfils2015segment}. The path between two middlepoints is computed by the shortest path algorithm of distributed protocols. Thus, this approach may mix requirement-contradictory flows over the same (shortest) path, which may violates the requirements on routing path (\emph{i.e.}, $f_1$ and $f_2$ in Fig. \ref{subfig:path}). To disperse traffic flows over different paths, this approach would introduce more middlepoint labels to each packet, bearing the same problems of \emph{hop encapsulation}. 

\noindent\textbf{Path encapsulation cannot always be scalable as expected.} 
Some proposes to utilize compression approach to pre-install all the desired paths and puts a global label in each packet header to specify which path should be applied \cite{chen2015xpath}.
This approach has limited scalability when applying to asymmetrical network topologies since the condition of its state compression algorithm does not be satisfied. 

\subsection{Explicit \brick\ is more powerful}
\begin{figure}
\centering
\includegraphics[width=0.9\linewidth]{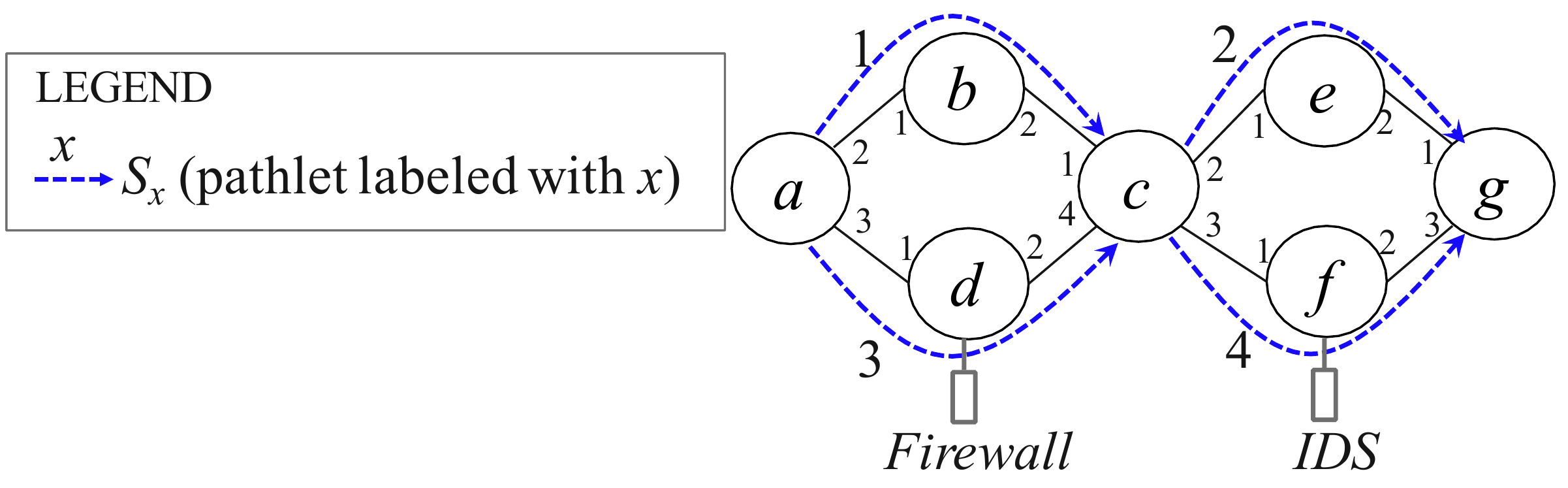}
\caption{An example of \bricks.
}\label{fig:segment-example}
\vspace{-2ex}
\end{figure}

The key intuition exploited by \sr\ is that we can profitably: (1) explicitly define \bricks/subpaths with the power of SDN, and (2) represent paths as concatenations of \bricks.  
In the example of Fig. \ref{fig:routemap}, for instance, \sr\ detects that all the desired paths overlap at \bricks\ $(a,b,c)$, $(c,e,f)$, $(a,d,c)$, $(c,g,f)$ (see Fig. \ref{fig:segment-example}). Moreover, \sr\ validates that 1) all the \bricks\ can be installed in the restricted flowtable, 2) and all the desired paths can be concatenated by \bricks. If applying a label to each \brick, \sr\ could represent those paths as concatenations of $P(f_1)=(1,2),P(f_2)=(1,4),P(f_3)=(3,2),P(f_4)=(3,4)$ respectively.

\noindent\textbf{\sr\ saves much rule space of flow table}.
When run on the example in Fig. \ref{fig:routemap}, \sr's rule overhead of installing \bricks\ in Fig. \ref{fig:segment-example} is much less than that of pushing all the paths to switches ($50\%$ rule saving on intermediate switches $b,d,c,e,f$ compared with Hop-by-Hop approach). 

\noindent\textbf{\sr\ hugely reduces the number of added labels}.
\sr's labels added in each packet of flows (say $f_i$, $i=1,..,4$) is only 2. This is much more efficient ($60\%$ label saving) than \emph{hop encapsulation} that would insert 5 labels into each packet.

\noindent\textbf{\sr\ efficiently preserves fine-grained path control}.
For the flows of different types, \sr\ puts their data over granular paths by stitching the explicitly defined \bricks\ together. Note that the suspicious traffic data (say flow $f_4$) would bypass the firewall if it adopts the \emph{middlepoint encapsulation} approach that would steer its traffic data over the shortest path ($a,b,c$) to reach the specified middlepoint (say $c$ if specified).  

\section{\sr\ Overview}
\label{sec:overview}
Fig. \ref{fig:overview} overviews \sr. We now describe the formal definition of \brick\ \sr\ adopts, the \sr's framework (section \ref{subsec:framework}) and the path construction process as well as packet forwarding process in the context of \sr. We use the terms switch and node interchangeably.

\noindent\textbf{\cbrick}. A \brick\ refers to a directed subpath $S=(v_0,v_1, \cdots, v_n)$ between node $v_0$ and $v_n$ in the controlled network, where $v_i$ denotes the network node and the nodes are all distinct. It is notice that either the start or the tail node can be the middle node, the ingress node or the egress node. 
Thus, a \brick\ differs much with the end-to-end path in that it is a part of the entire path.

In our \sr\ model, multiple \bricks\ are allowed between each node pair for the path flexibility/diversity purpose. Accordingly, there is a need to uniquely identify each \brick\ such that the overlapping nodes can perform correct actions instructed by \bricks. To this end, we assign a local \emph{\brick\ identifier} or $pid$ to each \brick. Each local identifier is only supported by the nodes over the associated \brick. The locality of $pid$ indicates that the same $pid$ can be assigned to multiple disjoint \bricks. Thus, we can use limited bits to express all the $pids$.
For convenience of description, we use $S_i$ to denote the \brick\ labeled with identifier $i$ in the following text, unless explicitly stated otherwise .

\subsection{Framework}\label{subsec:framework}
\begin{figure}
\centering
\includegraphics[width=0.65\linewidth]{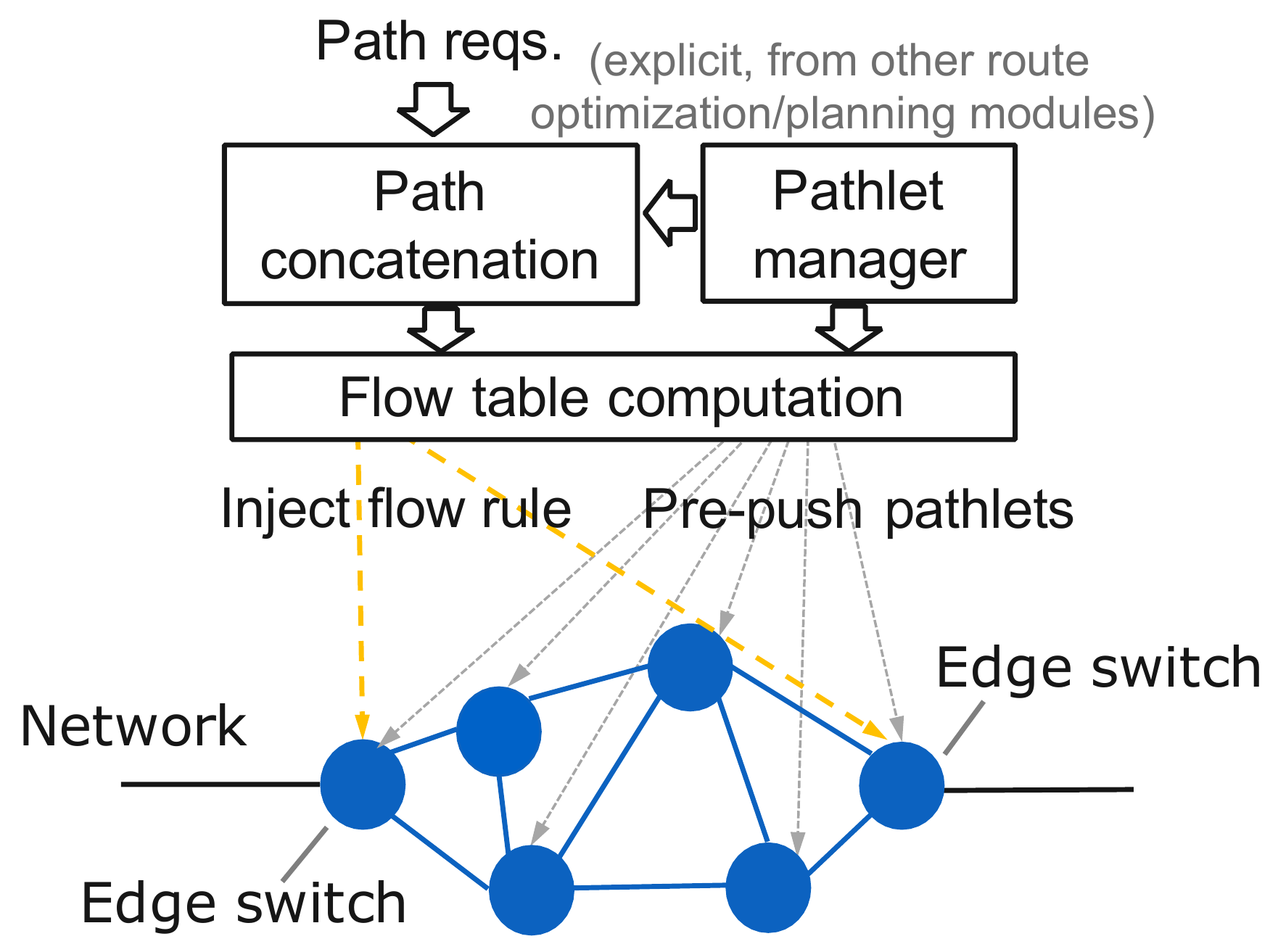}
\caption{Framework of \sr.
}\label{fig:overview}
\vspace{-2ex}
\end{figure}

Fig.\ref{fig:overview} overviews the framework of \sr. As with all the SDN-based approaches, we consider the network includes the controlled forwarding nodes and a logically centralized controller. The controller directly manages the network paths compliant with both the flow and network constraints. 

\emph{1) Functionality of a forwarding node:} In our framework, a controlled forwarding node shall be responsible for the following functionality.

\noindent\textbf{Edge node} provides the fine-grained path for each arriving packet/flow. In particular, the ingress node is responsible for encoding the desired path into the packet header as a list of \brick\ identifiers while the egress node would throw packets to corresponding outgoing interfaces.

\noindent\textbf{Core node} performs simple packet forwarding actions according to the \brick\ identifiers encapsulated in packets. Specifically, since the matching field is a \brick\ identifier, the look up operation is simple and easy.

\emph{2) Functionality of the controller:} Inside the controller, it has three main functional modules: \brick\ manager, path concatenation, and flow table computation. We outline the detail role of each below. 

\noindent\textbf{\cbrick\ manager} controls all the \bricks\ in the network, which plays a pivotal role in flexibly provisioning fine-grained paths for all differentiated network services and applications. In the implementation of \sr, \emph{pathlet manager} selects the \bricks\ that should be installed at the early stage of network configuration. Besides, this module provides the global perspective of \bricks\ for the \emph{path concatenation} module. (Section \ref{sec:segment_selection})

\noindent\textbf{Path concatenation} provides support for computing the \rpaths, taking the full end-to-end paths and \bricks\ as inputs. Indeed, this module focuses only on path concatenation, which is only a small subset of control plane functionality. It is important to clarify that \sr\ does none of additional route planning related tasks \cite{2013b4,2013swan,2015defo}. Considering this module supports the necessary functionality, we leave the route planning to other programs out of consideration for clean separation of network function modularity. Thus, the input only involves the explicit desired paths. For each explicit path request, this module efficiently transforms it into qualified \rpath\ through stitching minimum \bricks\ together. (Section \ref{sec:path_construction})

\noindent\textbf{Flow table computation} synthesizes rules for the \brick\ pre-installation and the flow path setup. We here describe the rules at a high-level and do not restrict the concrete implementation. 

For a \brick, the module crafts stateless rules $r(pid, action)$ for the middle network nodes to match against $pid$ and apply corresponding $action$ on matched packet. In particular, a \brick\ involves two types of actions: $forward(v,o)$ and $pop(v,o)$. A $forward(v,o)$ instructs a switch $v$ to put a matched packet on its outgoing interface $o$. While a $pop(v,o)$ instructs a switch $v$ firstly pop the top \brick\ identifier of a matched packet and then forward that packet to the outport $o$. 
In particular, for a \brick\ $S_i=(v_0,v_1, \cdots, v_n)$, \sr\ computes rules $r(i, forward(v_k,o))$, $(k=0,...,n-2)$ and a rule $r(i, pop(v_{n-1},o))$ for the nodes along that \brick. 

While for a requesting flow, this module crafts a stateful flow rule $r(fid, action)$ to instruct switches match against the flow identifier $fid$ (any specified $k-tuple$) and apply action ($insert$ or $forward$ ) on the packets belonging to the matched flow. The action $insert(v,f,sl,o)$ instructs switch $v$ firstly insert a list of \brick\ identifiers $sl$ to the packets of flow $f$ and then forward packets to outport $o$. Specifically, a flow rule $r(f, insert(v_{in},f,sl))$ should be installed at ingress switch $v_{in}$ and a rule $r(f, forward(u_{out},f,o))$ is designed to be placed on egress switch $u_{out}$. 

\noindent\textbf{What matters:} 
When talking about \sr\ 
, one would ask the following two questions: 1) how many and which \bricks\ should the network pre-install in the network such that all of the desired paths can be concatenated and the flow table capacity is enough, 2) how to concatenate path with \bricks\ such that the \brick\ identifiers or labels added on each packet is always limited.
In the below, the first problem is addressed by our optimization model and Lagrangian heuristic (Section \ref{sec:segment_selection}). The second issue is resolved by our efficient optimal path concatenation algorithm (Section \ref{sec:path_construction}).

\section{\cbrick\ Selection}
\label{sec:segment_selection}
This section details on how to select \bricks. We first state the \brick\ selection problem (section \ref{subsec:problem}), and then 
elaborate on the formulation of a linear programming for it (section \ref{subsec:formulation}). Finally, to obtain a selection solution within polynomial computational complexity, we show how to decompose problem based on Lagrangian relaxation (section \ref{subsec:dualproblem}, \ref{subsec:decomposition}) and present our \brick\ selection algorithm based on subgradient algorithm (section \ref{subsec:multiplier}, \ref{subsec:lagrangian-heuristic}).

\subsection{The problem description}\label{subsec:problem}
The \emph{\brick\ manager} decides what \bricks\ should be installed in the dataplane such that all the desired path can be concatenated by them.

\noindent\textbf{Input.}\label{P1:input}  
We assume an set of numerous desired paths $\mathcal{P}^{D}$ and a large set of \brick\ candidates $\mathcal{S}^C$. 
$\mathcal{P}^D$ can be computed by the logically centralized controller according to the history log/data, reservation services or optimization goals. Note that, \sr\ only admits different routing paths between each ingress-egress pair. When determining \bricks, one would try to deduce \bricks\ according to certain topological characteristics (\emph{e.g.}, overlapping, similarity) among all the desired paths at first thought. However, it is very difficult due to the enormous variations in paths and their topological relationships. To simplify the selection problem, we adopt a large set of \brick\ candidates $\mathcal{S}^C$ and try to select a small group \bricks\ from that big collection to concatenate all the desired paths. 

\noindent\textbf{Constraints to be meet} including flow table capacity constraint and length of path label constraint.

\emph{C1: Flow table constraint.} The flow table size is a constraint on each network node. A flow table contains entries that specify the forwarding rules for different \bricks, and its capacity is limited. To express this constraint quantitatively, we use $c_v^{free}$ to denote the free flow table capacity of node $v\in V$.

\emph{C2: Length of path label constraint.} The length of label to encode a path is another constraint. The length of path label equals to the number of \bricks\ in a \rpath\ and reflects the label overhead of the packet who applies that concatenation. Such packet overhead may waste precious bandwidth. It is notice that the path label optimization is done in our \emph{path concatenation module}. We here only constrain the maybe \brick\ number of a \rpath\ to a loose upper bound for the purpose of pre-optimization.
To meet this constraint, we adopt $m^{max}$ to represent the maximum allowed \bricks\ of \rpath.

\noindent\textbf{Output.}
The \emph{\brick\ manager module} returns the a small group of \bricks\ $\mathcal{S}^{select} \subseteq \mathcal{S}^C$ such that the \rpaths\ $\mathcal{P}^{concatenate} \subseteq \mathcal{P}^D$ is maximized. The best solution is to be $\mathcal{P}^{concatenate} = \mathcal{P}^D$, which is alway promised by our strategy algorithm as shown in section \ref{sec:evaluation}.

\subsection{The formulation}\label{subsec:formulation}
We model the network as a directed graph $G = (V, E)$, where set $V$ represents the network nodes and set $E$ models the directed links connecting nodes.
From the problem described in Section \ref{subsec:problem}, we can easily get the links and nodes that each path or \brick\ candidate go through. For each link $e \in E$, let $b_{e,P} = 1$ (resp. $b_{e,P} = 0$) if path $P$ traverses (resp. does not traverse) through link $e$ and $a_{e,S} = 1$ (resp. $a_{e,S} = 0$) if \brick\ candidate $S_i$ passes (resp. does not pass) link $e$. Besides, we let $l_P = \sum_{e\in E} b_{e,P}$ to denote the length of path $P$ and $l_S=\sum_{e\in E} a_{e,S}$ to denote that of \brick\ candidate $S$. Moreover, we let $d_{v,S}$ denote the rules needed on node $v$ to install \brick\ $S$. 

We formulate the \brick\ selection problem \textbf{P} as an \emph{Integer Linear Programming} (ILP). To formulate our problem, we introduce the following decision variables. We define a binary decision variable $x_{S,P}$ to be $1$ if \brick\ $S$ is a component of the concatenation of path $P$, and $0$ otherwise. 
for each \brick\ candidate in $\mathcal{S}^C$, we define the decision variable $t_{S} \in \{0,1\}$ to denote whether it is selected.
For each path in $\mathcal{P}^D$, we define the binary decision variable $y_P$ to be 1 if path $P$ cannot be concatenated and 0 otherwise.
To maximize the \rpaths\ (minimize the unconcatenated paths), now the problem \textbf{P} can be formulated as follows:
\begin{align}
& Z_{IP} = \min \sum_{P \in \mathcal{P}^D} y_{P} \label{ilp:obj}\\
\text{Subject to: } &\forall e\in E, P\in \mathcal{P}^D: \sum_{S \in \mathcal{S}^C} a_{e,S} \times x_{S,P} \le b_{e,P} \label{ilp:cons1}\\
&\forall P\in \mathcal{P}^D: l_P\times (1-y_P)= \sum_{S \in \mathcal{S}^C}  l_S\times x_{S,P}\label{ilp:cons2}\\
&\forall P\in \mathcal{P}^D: \sum_{S \in \mathcal{S}^C} x_{S,P} \le m^{max} \label{ilp:cons3}\\
&\forall S\in \mathcal{S}^C: \sum_{P \in \mathcal{P}^D}x_{S,P} \le |\mathcal{P}^D|\times t_S \label{ilp:cons4}\\
&\forall v\in V: \sum_{S \in \mathcal{S}^C}d_{v,S}\times t_S \le c_v^{free} \label{ilp:cons5} \\
& \forall S \in \mathcal{S}^C, \forall P \in \mathcal{P}^D: x_{S,P} \in \{0, 1\} \label{ilp:consx}\\
& \forall P \in \mathcal{P}^D: y_{P} \in \{0, 1\}  \label{ilp:consy}\\
& \forall S \in \mathcal{S}^C: t_{S} \in \{0, 1\}  \label{ilp:const} 
\end{align}

Constraints Ineq. \eqref{ilp:cons1} and Eq. \eqref{ilp:cons2} constrain that the concatenation of a desired path should have the same path structure with it. Constraints Ineq. \eqref{ilp:cons3} are length of path label/\brick\ constraints on each \rpath. Ineq. \eqref{ilp:cons4} and Ineq. \eqref{ilp:cons5} express the flow table constraints on each node.

After solving the above problem, for the desired path $\mathcal{P}^D$, we can obtain the maximum number of \rpaths\ $\mathcal{P}^{concatenate}$ (through $y$) and the \brick\ selection solution $\mathcal{S}^{select}$ (through $t$), under the constraints on flow table size and \rpath\ length. 

To solve the problem \textbf{P} within polynomial computational complexity, we design an \emph{\brick\ selection heuristic} (\alg) to obtain a near-optimal solution. Based on decomposition techniques of Lagrangian relaxation and ``divide and conquer'', subgradient algorithm, and branch-and-bound technique, the \alg\ can be efficiently implemented. The following shows more details.

\subsection{Lagrangian Dual Problem}\label{subsec:dualproblem} 
We dualize the constraints in Ineq. \eqref{ilp:cons4} to obtain the Lagrangian dual problem as the relaxing problem which can be further decomposed into two sub-problems, each can be solved within polynomial computational complexity.

In particular, we first relax the Ineq. \eqref{ilp:cons4} by bringing them to the objective function with associated Lagrangian multiplier vector $\bm{\lambda}=\{\lambda_S \ge0, S\in \mathcal{S}^C\}$. We get the Lagrangian relaxation problem $\textbf{P}_{LR}$ as follows:
\begin{equation}\label{ilp:relax-obj}
\begin{split}
 Z_{LR}(\bm{\lambda})&= \min \sum_{P \in \mathcal{P}^D}y_P+\sum_{S \in \mathcal{S}^C}\lambda_S(\sum_{P \in \mathcal{P}^D}x_{S,P}-|\mathcal{P}^D|t_S) \\
&= \min \sum_{P \in \mathcal{P}^D}(y_P+\sum_{S \in \mathcal{S}^C}\lambda_Sx_{S,P})-|\mathcal{P}^D|\sum_{S \in \mathcal{S}^C}\lambda_S t_S
\end{split}
\end{equation}
\begin{equation} \nonumber
\text{Subject to } \eqref{ilp:cons1}-\eqref{ilp:cons3} \text{ and } \eqref{ilp:cons5}-\eqref{ilp:const}.
\end{equation}

For any Lagrangian multiplier $\bm{\lambda}$, the value $Z_{LR}(\bm{\lambda})$ of the Lagrangian function is a lower bound on the optimal objective function value of the original minimization problem \textbf{P} \cite{fisher2004lagrangian}. Thus, to obtain the sharpest  possible lower bound, we would need to solve the following Lagrangian dual problem $\textbf{P}_{LD}$ of \textbf{P}:
\begin{equation} \label{ilp:dual}
Z_{LD} = \max_{\bm{\lambda} >0} Z_{LR}(\bm{\lambda})
\end{equation}

We observe that constraints Ineq. \eqref{ilp:cons1}-\eqref{ilp:cons3}, \eqref{ilp:consx}-\eqref{ilp:consy} only include the group of variables $\{x_{S,P}, y_P\}$ and Ineq. \eqref{ilp:cons5}, \eqref{ilp:const} are only related with the group of variables $\{t_S\}$. Therefore, problem $\textbf{P}_{LR}$ can be decomposed into two sub-problems involving clean separated variable set, $\textbf{P}_{Sub1}$ and $\textbf{P}_{Sub2}$ as follows,
\begin{equation}
Z_{Sub1}(\bm{\lambda}) = \min \sum_{P \in \mathcal{P}^D}(y_P+\sum_{S \in \mathcal{S}^C}\lambda_Sx_{S,P})
\end{equation}
under constraints Ineq. \eqref{ilp:cons1}-\eqref{ilp:cons3} and Ineq. \eqref{ilp:consx}-\eqref{ilp:consy}, and
\begin{equation}
Z_{Sub2}(\bm{\lambda}) = \min -|\mathcal{P}^D|\sum_{S\in \mathcal{S}^C}\lambda_St_S
\end{equation}
under constraints Ineq. \eqref{ilp:cons5} and Ineq. \eqref{ilp:const}.
 
We observe that problem $\textbf{P}_{Sub1}$ can be further decomposed into $|\mathcal{P}^D|$ mini-problems through ``divide and conquer'' approach. Finally, we can obtain the optimal solution for $\textbf{P}_{Sub1}$ by compositing the solutions of mini-problems.
While for the $\textbf{P}_{Sub2}$, it is the general 0-1 linear programming problem, which can be efficiently solved through the exact algorithm of branch-and-bound. In particular, we obtain the optimal solution of the $\textbf{P}_{Sub2}$ through the bound-and-bound algorithm in \cite{1990knapsack}.

Given $\bm{\lambda}$, we can compute $Z_{LR}(\bm{\lambda})$ through the optimal objective values of its two sub-problems. In particular, we obtain $Z_{LR}(\bm{\lambda}) = Z_{Sub1}(\bm{\lambda}) + Z_{Sub2}(\bm{\lambda})$.

\subsection{Decomposition for sub-problem $\textbf{P}_{Sub1}$}\label{subsec:decomposition}
We adopt the ``divide and conquer'' approach to further decompose the problem $\textbf{P}_{Sub1}$ into $|\mathcal{P}^D|$ independent mini-problems, each involves a single desired path. For a single path $P \in \mathcal{P}^D$, we only need to solve the following optimization problem $\textbf{P}_{Sub1,P}$.
\begin{align}
Z_{Sub1,j}&(\bm{\lambda})  = \min y_P+\sum_{S \in \mathcal{S}^C}\lambda_Sx_{S,P} \label{ilp:sub1Pj_obj}\\
\text{Subject to: } &\forall e\in E: \sum_{S \in \mathcal{S}^C} a_{e,S} \times x_{S,P} \le b_{e,P} \label{sub1Pj:cons1}\\
&l_P\times(1-y_P) = \sum_{S\in \mathcal{S}^C} l_S\times x_{S,P}\label{sub1Pj:cons2} \\
&\sum_{S \in \mathcal{S}^C}x_{S,P} \le m^{max} \label{sub1Pj:cons3} \\
& y_P \in \{0,1\}, \forall S\in \mathcal{S}^C: x_{S,P} \in \{0,1\} \label{sub1Pj:consxy}
\end{align}
 
This above problem can be efficiently solved by adopting the branch-and-bound technique.

After obtaining the optimal objective value $Z_{Sub1,P}(\bm{\lambda})$ of problem $\textbf{P}_{Sub1,P}, P\in \mathcal{P}^D$, we can easily compute $Z_{Sub1}^{\bm{\lambda}}=\sum_{P \in \mathcal{P}^D} Z_{Sub1,P}(\bm{\lambda})$.

\subsection{Multiplier selection and update}\label{subsec:multiplier}
It is clear that the selection of multipliers $\bm{\lambda}$ is important for obtaining a good quality or tightness lower bound of problem \textbf{P}. We start with an initial multiplier vector and adopt the subgradient optimization to iteratively update $\bm{\lambda}$ because of its well-known efficiency.

We choose a starting point $\bm{\lambda}^0 = 10^{-3}$. While at the $(k+1)th$ iteration, multiplier vector $\bm{\lambda}^{k+1}$ can be obtained by $\bm{\lambda}^{k+1}=max\{\bm{\lambda}^{k}+\theta_k\bm{\mu}^{k}, 0\}$, in which $\bm{\lambda}^{k}$, $\theta_k$, $\bm{\mu}^k$ respectively denotes the multiplier vector, the step size, the subgradient vector used in the $kth$ iteration. Their detail calculations are as follows:
\begin{enumerate}[-]
\item $\bm{\mu}^k$: The subgradient vector $\bm{\mu}^k$ is computed by $\mu^k_S=\sum_{P\in \mathcal{P}^D}x^k_{S,P} - |\mathcal{P}^D|t^k_S,\ S\in \mathcal{S}^C$, where $x^k_{S,P}$ and $t^k_S$ respectively denotes the values of variables $x_{S,P}$ and $t_S$ in the optimal solution of problem $\textbf{P}^{\bm{\lambda}}$ obtained at $\bm{\lambda}^k$. 
\item $\theta_k$:  The positive step size $\theta_k$ can be acquired by a common method in \cite{fisher2004lagrangian} as follows: $\theta_k=\frac{\beta_k(z_{UP}^k-z_{LB}^k)}{||\bm{\mu}^k||^2}$, where $\beta_k$ is a positive scalar satisfying $0< \beta_k \le2$ and $z_{UP}^k$ (resp. $z_{LB}^k$) denotes an upper (resp. lower) bound on the optimal objective of the original optimization programming at iteration $k$.
To obtain a lower bound closer to the optimal value, we compute $z_{LB}^k$ by $max\{B^{\bm{\lambda}^k}, z_{LB}^{k-1}\}$, where $B^{\bm{\lambda}^k}$ is the objective value of the Lagrangian relaxation programming at $\bm{\lambda}^k$.
Besides, the objective value of a feasible solution of the original problem is obviously an upper bound on the optimal objective of the primal problem. Thus, to obtain the smaller upper bound, the $z_{UP}^k$ is computed by $min\{z_{FE}^k,z_{UP}^{k-1}\}$, where $z_{FE}^k$ respectively denotes the objective value of a feasible solution at iteration $k$. 

\item $z_{FE}^k$:
To construct a primal feasible solution $z_{FE}^k$, we let $\bm{t}^k = \{t^k_S, S\in \mathcal{S}^C\}$ be known parameters and solve the optimization programming with only constraints Ineq. \eqref{ilp:cons1}-Ineq. \eqref{ilp:cons4}. It is clearly that the value of $\bm{t}^k$ have already satisfied constraints Ineq. \eqref{ilp:cons5}. Therefore, the obtained solution obviously satisfies the feasibility.
\end{enumerate}

\subsection{\cbrick\ selection heuristic (\alg) based on Lagrangian relaxation}\label{subsec:lagrangian-heuristic}
We now describe our Algorithm \ref{alg:lagrangian} which is designed based on Lagrangian relaxation and subgradient optimization to solve the original programming. Specifically, sub-problems $\textbf{P}_{Sub1}$ and $\textbf{P}_{Sub2}$ are solved with multiplier vector $\bm{\lambda}^k$ at iteration $k$. At each iteration $k$, we can obtain a feasible solution for the original problem based on $\bm{t}^k$ and an upper bound of the original programming. We maintain an upper bound $z^{k}_{UP}$ as the smallest upper bound we have obtained within $k$ iterations. On the other hand, $z^k_{LB}$ denotes the maximum value of the objective of problem $\textbf{P}_{LR}$ after $k$ iterations, which is a lower bound the the original programming.

To obtain a satisfactory solution within limited computation time, the Algorithm \ref{alg:lagrangian} is stopped when one of the conditions is satisfied: 1) the number of iteration $k$ reaches the iteration limit $|\mathcal{P}|$; 2) the difference between $z^k_{LB}$ and $z^k_{UP}$ is less than a threshold $\epsilon^{*}$; 3) the lower bound does not increase for more than a number of iterations $T'$. After the algorithm is terminated, it returns the feasible solution $\pi^{*}$, which reaches to the minimum upper bound during the iterations.
\begin{algorithm}
\caption{OneRoundSelection($\mathcal{S}^C$, $\mathcal{P}^D$)}\label{alg:lagrangian}
    \begin{algorithmic}[1] 
            \State $k\gets 1$ and $t'\gets 0$;
            \State $z^{0}_{UP}\gets +\infty$ and $z^{0}_{LB}\gets -\infty$;
            $\beta_1 \gets 2$ and $\epsilon^{1} \gets +\infty$
            \State Let $\lambda^1_S \gets 10^{-3}, S \in \mathcal{S}^C$;
            \While{$k<|\mathcal{P}^D|$ \textbf{and} $\epsilon^t>\epsilon^{*}$ \textbf{and} $t'<T'$}
                \State Solve problem $\textbf{P}_{Sub1}$; Obtain $Z_{Sub1}^{k}$ and $x^k_{S,P}, S\in \mathcal{S}^C, P\in \mathcal{P}^D$;
				\State Solve problem $\textbf{P}_{Sub2}$; Obtain $Z_{Sub2}^{k}$ and $t^k_S, S\in \mathcal{S}^C$;

             	\State $Z_{LR}^k = Z_{Sub1}^{k} + Z_{Sub2}^{k}$;
             	\State Let \mbox{\boldmath $t^k$} be an given parameters and solve the
             	 problem \textbf{P} with equations \eqref{ilp:cons1}-\eqref{ilp:cons4}, let $z_{FE}^k$ be the objective value of the solving problem;
             	\State Let $z^k_{UP} = min\{z_{FE}^k, z^{k-1}_{UP}\}$;

             	\If{$z^t_{UP} < z^{k-1}_{UP}$} 
             		\State Record the feasible solution $R^{*}$;
         		\EndIf

         		\State Let $z^k_{LB} = max\{Z^k_{LR},z^{k-1}_{LB}\}$;

         		\If{$z^k_{LB} > z^{k-1}_{LB}$}
         			\State $t' = 0$;
     			\Else
     				\State $t' = t'+1$;
 				\EndIf

 				\If{$t' \geq 4$}
 					\State $\beta_k = \beta_{k-1}/2$;
				\EndIf

				\State $\epsilon^{k+1} = z^k_{UP}-z^k_{LB}$;
				\State // update the multipliers $\bm{\lambda}$
				\State $\mu^k_S =\sum_{P \in \mathcal{P}^D} x^k_{S,P} - |\mathcal{P}^D|t^k_S, S\in \mathcal{S}^C$;
				\State $\theta^k = \frac{\beta_k(z^k_{UP}-z^k_{LB})}{\vert\vert \bm{\mu}^k \vert\vert^2}$;
				\State $\lambda^{k+1}_S = max\{\lambda^k_S + \theta^k\mu^k_S,0\}, S\in \mathcal{S}^C$;
				\State $k = k+1$;
            \EndWhile
            \State \Return{$R^{*}$ and $z^{k-1}_{UP}$};

    \end{algorithmic}
\end{algorithm}

In very rare cases, some of the desired paths cannot be concatenated after running the \alg\ only once. This is because of the improper input of \brick\ candidates $\mathcal{S}^C$. To maximize the \rpaths, we present algorithm \ref{alg:pathletselect} that runs \alg\ repeatedly until all the paths can be concatenated (see Line \ref{line:beginwhile}-\ref{line:endwhile} in Algorithm \ref{alg:pathletselect}). Specifically, at each running time, \alg\ takes only the unconcatenated paths and a set of random \brick\ candidates as input.
\begin{algorithm}
\caption{Select\cbrick($\mathcal{P}^{D}$)}\label{alg:pathletselect}
\begin{algorithmic}[1]
	\State // $\mathcal{P}^D:$ all the desired paths
	\State $\mathcal{S}^{select}\gets \varnothing$; \Comment{The selected \bricks}
	 \State $\mathcal{P}^{concatenate}\gets \varnothing$;\Comment{The \rpaths}
	 \State $n\gets 1 $;
	\While{$n\leq N$}\Comment{$N$ is the iteration limit} \label{line:beginwhile}
		\State $\mathcal{P}^{non} = \mathcal{P}^D-\mathcal{P}^{concatenate}$; 
		\State Generate \brick\ candidates $\mathcal{S}^C$; \Comment{Randomly}
		\State $R^{*}$ = \Call{Selection}{$\mathcal{S}^C$, $\mathcal{P}^{non}$};
		\State \text{Extract the \rpaths\ $\mathcal{P}_n$ and the selected }
		\Statex \text{\hspace{1em} \bricks\ $\mathcal{S}_n$ from $R^{*}$;}
		\State $\mathcal{P}^{concatenate}$.addPath($\mathcal{P}_n$);
		\State $\mathcal{S}^{select}$.add\cbrick($\mathcal{S}_n$);
		\If{$\mathcal{P}^{concatenate} = \mathcal{P}^D$}
			\State \textbf{return} $\mathcal{S}^{select}$;
		\EndIf
		\State Update $c_v^{free}, v\in V$ according to $\mathcal{S}^{select}$;
		\State $n\gets n+1$;
	\EndWhile\label{line:endwhile}
	\State \textbf{return} $\mathcal{S}^{select}$;
\end{algorithmic}
\end{algorithm}

\section{Path Concatenation}\label{sec:path_construction}
As described in section \ref{subsec:framework}, the task of \emph{path concatenation module} is to concatenate each on-demand path with the \bricks\ selected by \emph{segment manager module}.

\subsection{Problem description}
As we have described previous, as the path is encoded in each packet, the path label overhead should be minimized for saving bandwidth resource. Therefore, for a on-demand path $P$, to minimize the path label overhead, the problem here is to compute the minimum \bricks\ from the selected \bricks\ $\mathcal{S}^{select}$ to construct the concatenation for it.
We say a concatenation solution is optimal for a path when it has the minimum \bricks.
Although the \brick\ selection algorithm in section \ref{sec:segment_selection} has bounded the number of \bricks\ of a \rpath, it is a loose constraint and does not promises the optimal concatenation for each single path. 

\subsection{Problem analysis}
Since there is no competition for network resources when actually concatenating the requesting paths, we focus on the problem of processing one on-demand path. If there are  multiple path requests, we can compute their \rpath\ in parallel.
In particular, we exploit the fact that computing the optimal solution for a single path is not NP-hard: 
\begin{mytheorem}{An optimal concatenation for a given request $P$ can be computed in polynomial time. 

\noindent{Proof.} \emph{The proof is constructive. We first prune all \bricks\ $S\in \mathcal{S}^{select}$ whose traversing links do not be traversed by $P$. Adopting the above pruning method, the precise \brick\ candidates $\mathcal{S}^{P}$ for $P$ is obtained, which has ben proved to be $|\mathcal{S}^{P}|\le (\frac{l_P^2+l_P-2}{2})$. For each $S \in \mathcal{S}^{P}$, since $P$ goes through all links $e$ on it, thus it is a proper subset of $P$. Moreover, the number of \bricks\ used to concatenate $P$ must be less than $l_P$. Therefore, we can compute all the possible combinations $\mathcal{P}^{enumerate}$ ($|\mathcal{P}^{enumerate}|=\sum_{i=2}^{l_P-1}C^i_{\mathcal{S}^{P}}$) of \bricks\ in $\mathcal{S}^P$.
Since the number of possible combinations is limited, we can obtain the optimal one in polynomial time even adopting the enumeration approach, proving the theorem. 
}}
\end{mytheorem}

Theorem 1 is an important instruction for our path concatenation algorithm design, as it allows us to devise algorithm that always returns optimal concatenation.
However, the natural enumeration approach that verifies all the possible combinations is not efficient.

\subsection{Algorithm} 
We present an efficient path concatenation algorithm (\algpath) that strategically combines incremental enumeration with ``First-Fit'' to obtain the optimal \rpath\ (see Algorithm \ref{alg:path_construction}). Firstly, given the selected \bricks\ $\mathcal{S}^{select}$ and a requesting path $P$, \algpath\ extracts the precise \brick\ candidates for $P$. Then, \algpath\ performs the following three steps to compute the optimal \rpath\ for $P$.
Step 1: \emph{Enumerate} all the combinations with $m$ \bricks, where $m$ is initialized with 2 (Line \ref{alg2:enumeration}).
Step 2: \emph{Verify} the enumerated combinations obtained from step 1 one by one until finding a satisfactory one (Line \ref{alg2:checkbegin}-\ref{alg2:checkend}). If this step can not find a satisfactory \rpath\ under the current $m$, it will then increase the $m$ by $1$ and go to step 1. \algpath\ repeats the step 1, 2 until obtaining a feasible \rpath\ $P^{con}_{optimal}$ or reaching the possible maximum value of $m$ ($l_P$). 
Step 3: \emph{Nesting} the obtained $P^{con}_{optimal}$ if the \bricks\ included in it exceed the allowed maximum value $m^{max}$ (Section \ref{subsec:nesting}). Otherwise, \algpath\ returns the optimal solution $P^{con}_{optimal}$.

\begin{algorithm}[!htb]
\caption{ConstructPath($\mathcal{S}^{select},P$)} \label{alg:path_construction}
    \begin{algorithmic}[1] 
		    \State Extract the \brick\ candidates $\mathcal{S}^P$ from selected \bricks\ $\mathcal{S}^{select}$ for desired path $P$; \label{alg2:seg_extraction}
		    \State $P^{con}_{optimal}\gets \varnothing$; \Comment{Store the optimal \rpath.}
		    \State $m\gets 2$; \label{alg2:start}
		    \While{$m \le l_P$ and $P^{con}_{optimal} = \varnothing$}
		    	\State $\mathcal{P}^{enumerate}\gets Enumeration(\mathcal{S}^P, m)$; \label{alg2:enumeration}
		    	\For{$P^{'} \in \mathcal{P}^{enumerate}$} \label{alg2:checkbegin}
		    		\If{$P^{'} = P$}
		    			\State $P^{con}_{optimal} = P^{'}$, Break;
	                \EndIf
	            \EndFor \label{alg2:checkend}
		    	\State $m \gets m+1$;
		    \EndWhile
		    \If{$m \le m^{max}$} 
		    	\State return $P^{con}_{optimal}$;
	    	\Else
		    	\State $P^{con}_{optimal} \gets Nesting(P^{con}_{optimal})$ \label{alg2:shorten}, and return $P^{con}_{optimal}$;
	    	\EndIf
    \end{algorithmic}
\end{algorithm}

\subsection{\cbrick\ nesting scheme}\label{subsec:nesting}
In rare cases, a concatenation would still include many \bricks. This is because the tasks of \brick\ selection and concatenation computation are done separately. To handle such case, we adopt a nesting scheme to let each packet carry limited \brick\ identifiers at each hop switch. 
The main idea is simple. For long concatenations, we design representative \brick\ to denote several \bricks. The detail of a representative \brick\ is invisible at irrelevant nodes and only to be unfolded at the start point of that representative \brick.

\begin{figure}
\centering
\includegraphics[width=0.85\linewidth]{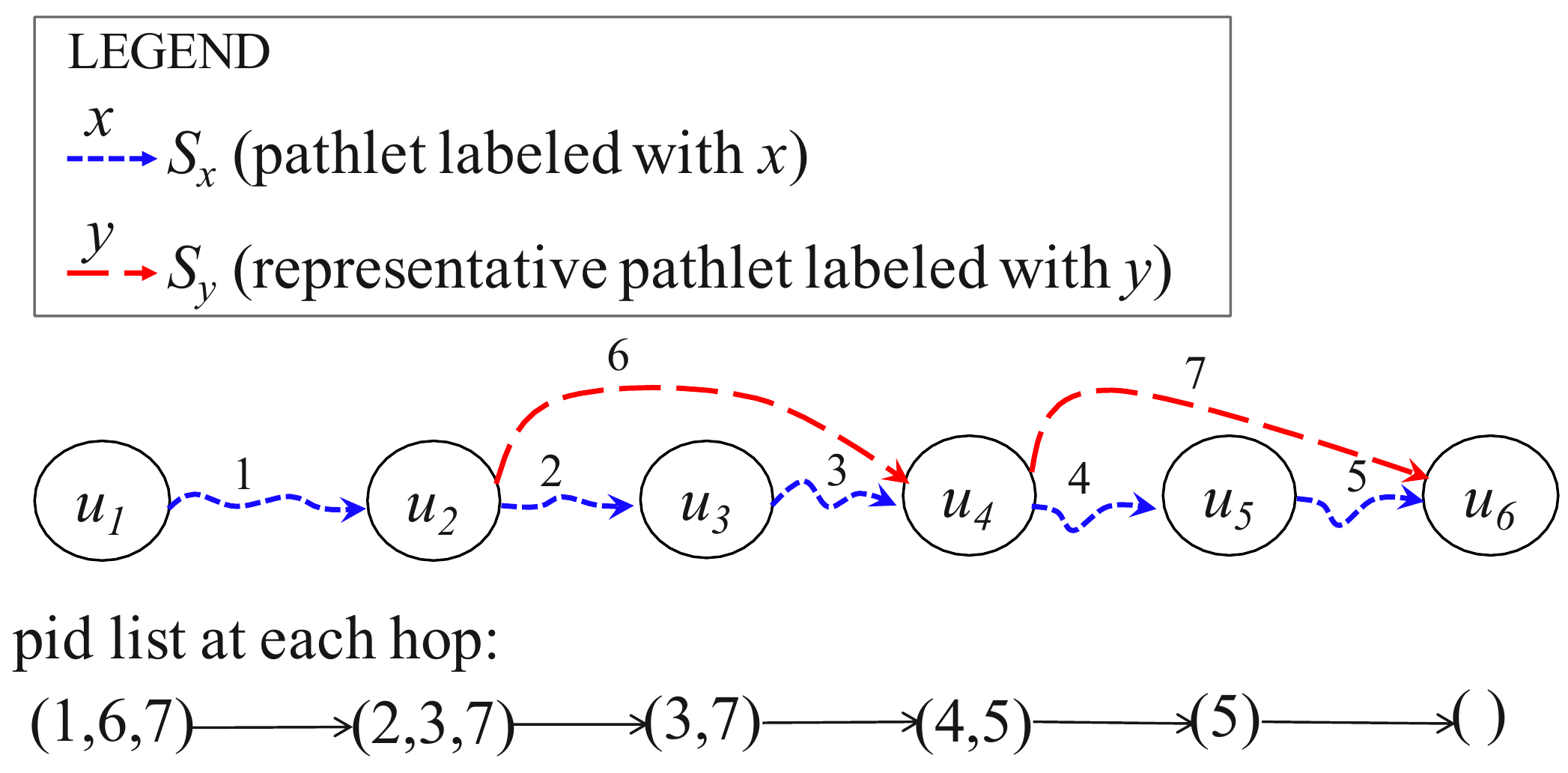}
\caption{An example of shortening a long \rpath.
}\label{fig:shorten-labels}
\vspace{-2ex}
\end{figure}

Consider the concatenation of a path $P_{u_1,u_6}$ is $\{S_1, S_2, S_3, S_4, S_5\}$ in Fig. \ref{fig:shorten-labels}. As stated in Section \ref{sec:overview}, each packet employing $P_{u_1,u_6}$ should carry five labels $sl=\{1,2,3,4,5\}$ at ingress switch $u_1$.  
The labels can be reduced to three if adopting two representive \bricks: $S_6$ (for \bricks\ $S_2$ and $S_3$) and $S_7$ (for \bricks\ $S_4$ and $S_5$). 
More generally, \sr\ can always compute a concatenation with very limited \bricks\ and \brick\ representatives for each path. 
\vspace{-1em}

\section{Performance evaluation}\label{sec:evaluation}
We evaluate \sr\ by performing 5,000 simulation experiments. In each experiment, we generate a large-scale of desired paths and run \sr\ on it. Besides, we collect the rules used to install the \bricks\ selected by \sr, the paths that can be concatenated by those selected \bricks\ and the length of path label of each path.
\begin{table}[!htb]
\vspace{-3ex}
\centering
\caption{Experimental Topologies}\label{table:topo-size}\vspace{0.8ex}
\tabcolsep=0.5em
{
\begin{tabular}{|c|ccccc|}

\hline
Topology name & rf3967 & rf1755 & rf1221 & rf6164 & rf3257 \\  \hline
 Nodes & 79 & 87 & 104 & 138 & 161 \\
Links  & 294 & 322 & 302 & 744 & 656 \\
\hline
\end{tabular}}
\vspace{-2ex}
\end{table}

As dataset, we adopt the inferred topologies from the Rocketfuel project \cite{sigcomm02-rocketfuel}. To generate large-scale desired paths, we assume four type of flows (protected, suspicious, bulk and time-sensitive) and compute policy path for each type of flows. In particular, we compute two disjoint paths for a protected flow, a path passing through a random waypoint for a suspicious flow, a bandwidth-sufficient path for a bulk flow and a low-latency path for a time-sensitive flow. In each experiment, we generate $m$ flows, each of random type, between each node pair and take the flow paths between all node pairs as the desired paths. To be fair, we only admit different paths between the same node pair. As for the \brick\ candidates, we generate $k$-simple paths between every node pair to make that. 
\vspace{-1.5ex} 
\begin{table}[!htb]
\centering
\caption{Rule overhead of \sr\ and that of Hop-by-Hop technique}\label{table:rule_saved}
\vspace{0.5ex}
\tabcolsep=0.6em
\scalebox{0.95}[0.9]
{
\begin{tabular}{|l|r|r|r|r|c|c|}
\hline
\multirow{2}{*}{ISPs} & \multirow{2}{*}{Paths} & \multicolumn{2}{c|}{Max.rules} & \multicolumn{2}{c|}{Avg.rules} &\multirow{2}{*}{$R_{Avgsave}$} \\ \cline{3-6}
 & & Hop-by-Hop & \sr\ & Hop-by-Hop & \sr & \\ 
 \hline
rf3967 & 26,636  & 9,281  & 467 & 2,538 & 343 & 86.48\% \\
rf1755 & 32,373  & 15,928 & 613 & 2,982 & 268 & 90.89\% \\
rf1221 & 41,132  & 21,970 & 641 & 3,095 & 209 & 93.25\% \\
rf6164 & 87,594  & 31,677 & 742 & 4,516 & 370 & 91.81\% \\
rf3257 & 109,516 & 45,316 & 926 & 4,872 & 286 & 94.13\% \\
\hline
\end{tabular}}
\end{table}

\begin{figure}
\centering
\includegraphics[width=0.6\linewidth]{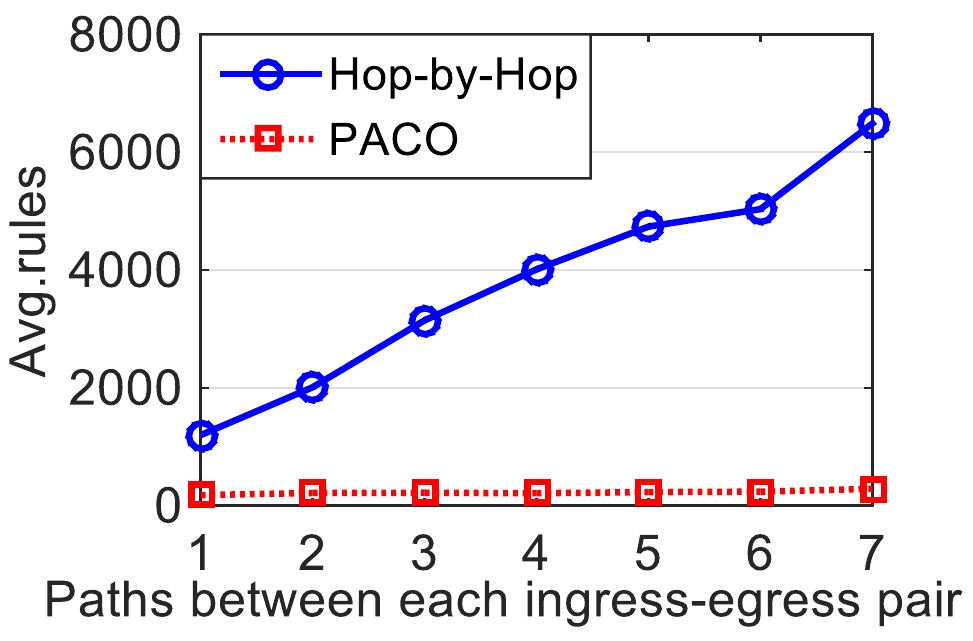}
\caption{The rules needed by \sr\ do not increase with the number of paths.
}\label{fig:rules}
\vspace{-1em}
\end{figure}

\noindent\textbf{\sr\ is scalable.} \sr\ hugely reduces the number of rules compared with Hop-by-Hop/OpenFlow-SDN technique. Table \ref{table:rule_saved} shows the results of experiments over a scenario there are four flows between each node pair. The results show \sr\ can save up to about $94\%$ average rules that would used by Hop-by-Hop technique. 
More interesting, the maximum rule number on a node are always less than $1000$, which indicates all of the \bricks\ selected by \sr\ could be installed in the commodity switch \footnote{The commodity switches have at least 2K capacity.}.
Besides, we rerun \sr\ over scenarios with increasing paths. The results of experiments on topology rf1221 depicted in Fig. \ref{fig:rules} show the average number of rules of \sr\ is always less than $300$ but that of Hop-by-Hop increases linearly with the paths.
This result indicates that \emph{\sr\ is scalable as the required rules would not increase with the number of paths.}

\begin{figure}
\centering
\vspace{-0.3em}
\includegraphics[width=0.6\linewidth]{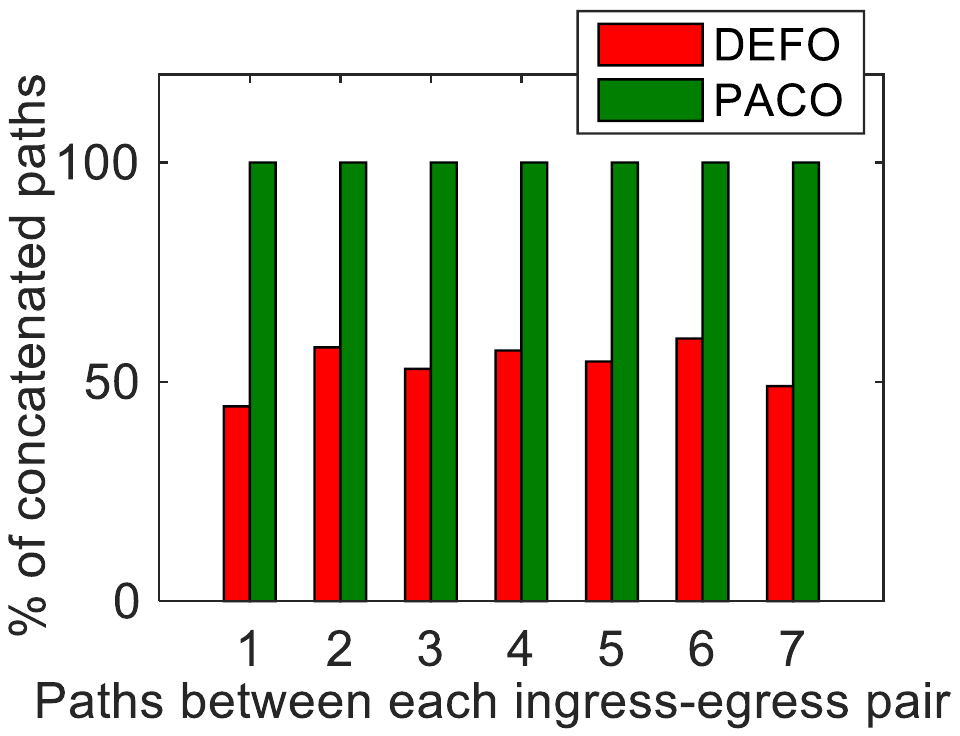}
\caption{The concatenated paths of \sr\ and of DEFO \cite{2015defo}.
}\label{fig:success}
\vspace{-1em}
\end{figure}

\noindent\textbf{\sr\ always concatenates all of the desired paths flexibly} with the \bricks\ selected by our algorithm in each and every experiment. 
We fix the maximum number of \bricks\ used to concatenate a path to three, as \cite{2015defo} states it can get significant TE improvement with that setting. We respectively run DEFO and \sr\ over the same set of desire paths. Results are displayed in Fig. \ref{fig:success}. \emph{\sr's $100\%$ concatenation indicates the preserving of path flexibility and marks an important difference with previous techniques based on middlepoint encapsulation.} The results show that DEFO cannot concatenate more than $\approx 60\%$ desired paths of the experiments on any topology. In contrast, \sr\ computes concatenations for all the desired paths in our experiments. This is because \sr\ adopts much  flexible \bricks\ between two middlepoints rather than shortest paths. The results of DEFO indicate that a set of shortest paths might fit well for a specific goal (\emph{i.e.}, TE) but would struggle to reach other goals (\emph{i.e.}, service function enforcement). 

\begin{figure}
\centering
\includegraphics[width=0.7\linewidth]{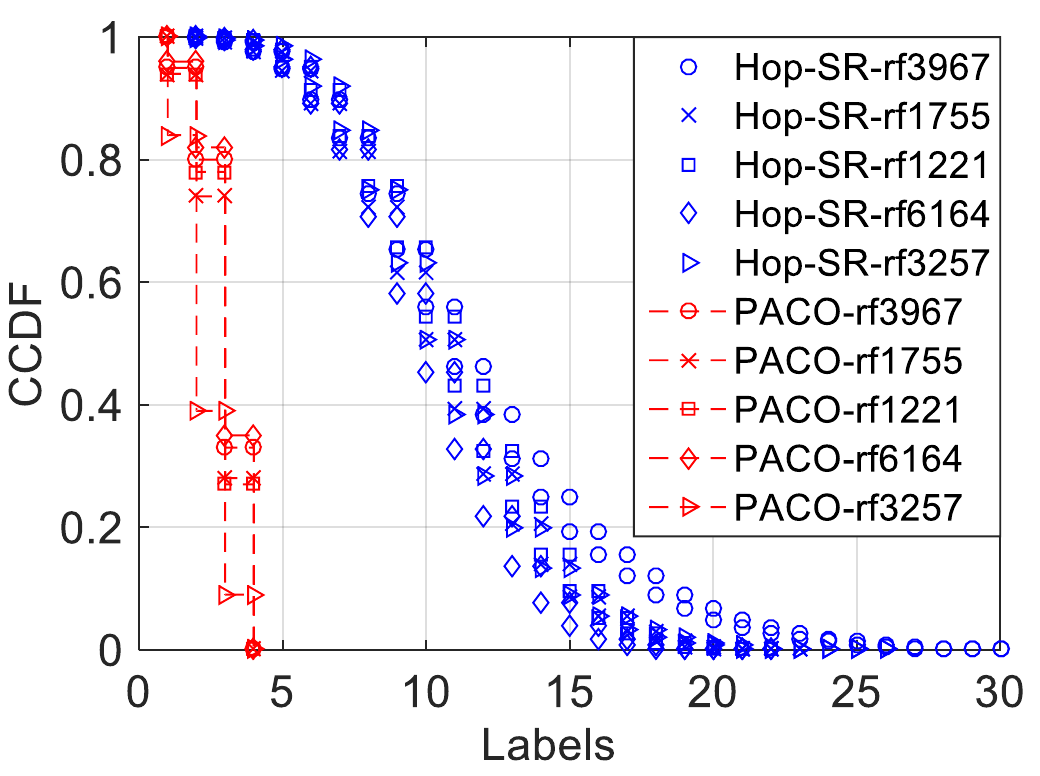}
\caption{The length of path label.
}\label{fig:labels}
\vspace{-1.2em}
\end{figure}

\noindent\textbf{\sr\ only needs to adopt limited labels} to encode a path. We evaluate the result of path labels for each experiment of each topology. Fig. \ref{fig:labels} depicts the Complementary CDF of the results. A data point $(x,y)$ in the figure indicates that for a fraction of $y$ of the desired paths need at least $x$ path labels. The results highlight that the \sr\ can concatenate all the paths with at most 4 path labels ($x=5, y=0$) across experiments on every topology. While the Hop-SR/encapsulation approach needs at least 10 path labels ($x=10$) to encode about $45\%$ paths ($y=0.45$). Indeed, for the Hop-SR method, the number of path labels increases linearly with the path length. \emph{The results indicate that \sr\ is efficient that the label overhead of all the paths is always lightweight}, while that of Hop-SR/encapsulation is heavy.

\section{Conclusion}\label{sec:conclusion}
In this paper, we present \sr, an Source Routing (SR) based framework to provide scalable and fine-grained path control for SDN networks. In the design, \sr\ would pre-install \bricks\ in the network and represents each on-demand path as a concatenation of \bricks. Each packet adopts the forwarding technique of SR. To reduce the label overhead SR brings while preserving path flexibility, we first design and implement a Lagrangian heuristic to determine which \bricks\ should be installed in the small flow table such that the all the fine-grained paths can be concatenated. And then we design optimal algorithm to computes the minimum \bricks\ to represent each path efficiently. The results of evaluation show that \sr\ outperforms previous approaches: In our experiments, it saves more than $94\%$ rules needed by Hop-by-Hop technique with very limited label overhead and supports $40\%$ more fine-grained paths than DEFO.

\bibliographystyle{IEEEtran}
\bibliography{paco}
\end{document}